\documentclass[aip,apl,reprint,superscriptaddress]{revtex4-1}

\usepackage{graphics,amsmath}
\usepackage{graphicx}
\usepackage{textcomp}
\usepackage{color}
\usepackage{ulem}

\begin{document}


\title{Quantitative disentanglement of coherent and incoherent laser-induced surface deformations by time-resolved x-ray reflectivity} 
%
%
%
%
\author{M.~Sander}
\author{J.-E.~Pudell}
\author{M.~Herzog}
\affiliation{Institut f\"ur Physik und Astronomie, Universit\"at Potsdam, Karl-Liebknecht-Str.\ 24-25, 14476 Potsdam, Germany}
\author{M.~Bargheer}
\affiliation{Institut f\"ur Physik und Astronomie, Universit\"at Potsdam, Karl-Liebknecht-Str.\ 24-25, 14476 Potsdam, Germany}
\affiliation{Helmholtz-Zentrum Berlin f\"ur Materialien und Energie GmbH, Wilhelm-Conrad-R\"ontgen Campus, BESSY II, Albert-Einstein-Str.\ 15, 12489 Berlin Germany}
\author{R.~Bauer}
\affiliation{Institut f\"ur Nanostruktur und Festk\"orperphysik, Universit\"at Hamburg, Jungiusstr.\ 11c, 20355, Germany}
\author{V. Besse}
\affiliation{IMMM CNRS 6283, Universit\'e du Maine, 72085 Le Mans cedex, France}
\author{V. Temnov}
\affiliation{Institute of molecules and materials of Le Mans, CNRS UMR 6283, 72085 Le Mans, France}
\author{P.~Gaal}
\affiliation{Institut f\"ur Nanostruktur und Festk\"orperphysik, Universit\"at Hamburg, Jungiusstr.\ 11c, 20355, Germany}
\email{pgaal@physnet.uni-hamburg.de}

\begin{abstract}
We present time-resolved x-ray reflectivity measurements on laser excited coherent and incoherent surface deformations of thin metallic films. Based on a kinematical diffraction model, we derive the surface amplitude from the diffracted x-ray intensity and resolve transient surface excursions with sub-\AA{} spatial precision and 70\,ps temporal resolution. The novel analysis allows for decomposition of the surface amplitude into multiple coherent acoustic modes and a substantial contribution from incoherent phonons which constitute the sample heating.
\end{abstract}

\maketitle

Ultrafast photoacoustics\cite{Ruel2015a} has become an established method to probe the interaction of optical\cite{Weiss2014a}, electronic\cite{Wang2016a} and magnetic\cite{kim2012a} properties with the crystal lattice in solids. It employs strain pulses that are generated by absorption of femtosecond light pulses in an optoacoustic  transducer.\cite{Grah1986a} Subsequent lattice dynamics can be probed either optically or by ultrafast x-ray diffraction.\cite{Rose1999a} Nowadays tailored longitudinal strain waves can be generated and monitored using time-resolved optical and x-ray techniques.\cite{Ruel2015a, herz2012c, Sand2017a} Mode selective excitation of coherent acoustic surface modes can be achieved with a Transient Grating (TG) technique\cite{Roge2000a}. In addition to Rayleigh-like Surface Acoustic Waves (SAW) this method also excites so-called Surface Skimming Longitudinal Waves (SSLW)\cite{Janu2016a,Janu2016b}. However, in any photoacoustic experiment, the main fraction of the deposited optical energy is stored in incoherent phonon excitations\cite{herz2012b, shay2011a}. The absolute magnitude of the coherent and incoherent excitation is hard to determine from purely optical experiments.
In this paper we perform a full decomposition of optically excited coherent acoustic surface and longitudinal waves which propagate with their respective group velocities and the concomitant thermal phonons which move only by diffusion. Our method allows for measuring the absolute deformation of a solid surface using time-resolved x-ray reflectivity (TR-XRR). This new method can resolve surface deformation with sub-\AA~spatial and 70\,ps temporal resolution.

The experiments were performed at the ID09 beamline of the European Synchrotron Radiation Facility (ESRF) in Grenoble, France. The beamline is equipped with a commercial Ti:Sapphire laser amplifier (Coherent Legend) which delivers 800\,nm optical pulses with a duration of 600\,fs at a repetition rate of 1\,kHz. The laser is synchronized to the storage ring to allow for tuning the pump-probe delay with a precision of better than 5\,ps.

\begin{figure}
  \centering
  \includegraphics[width = 0.8\columnwidth]{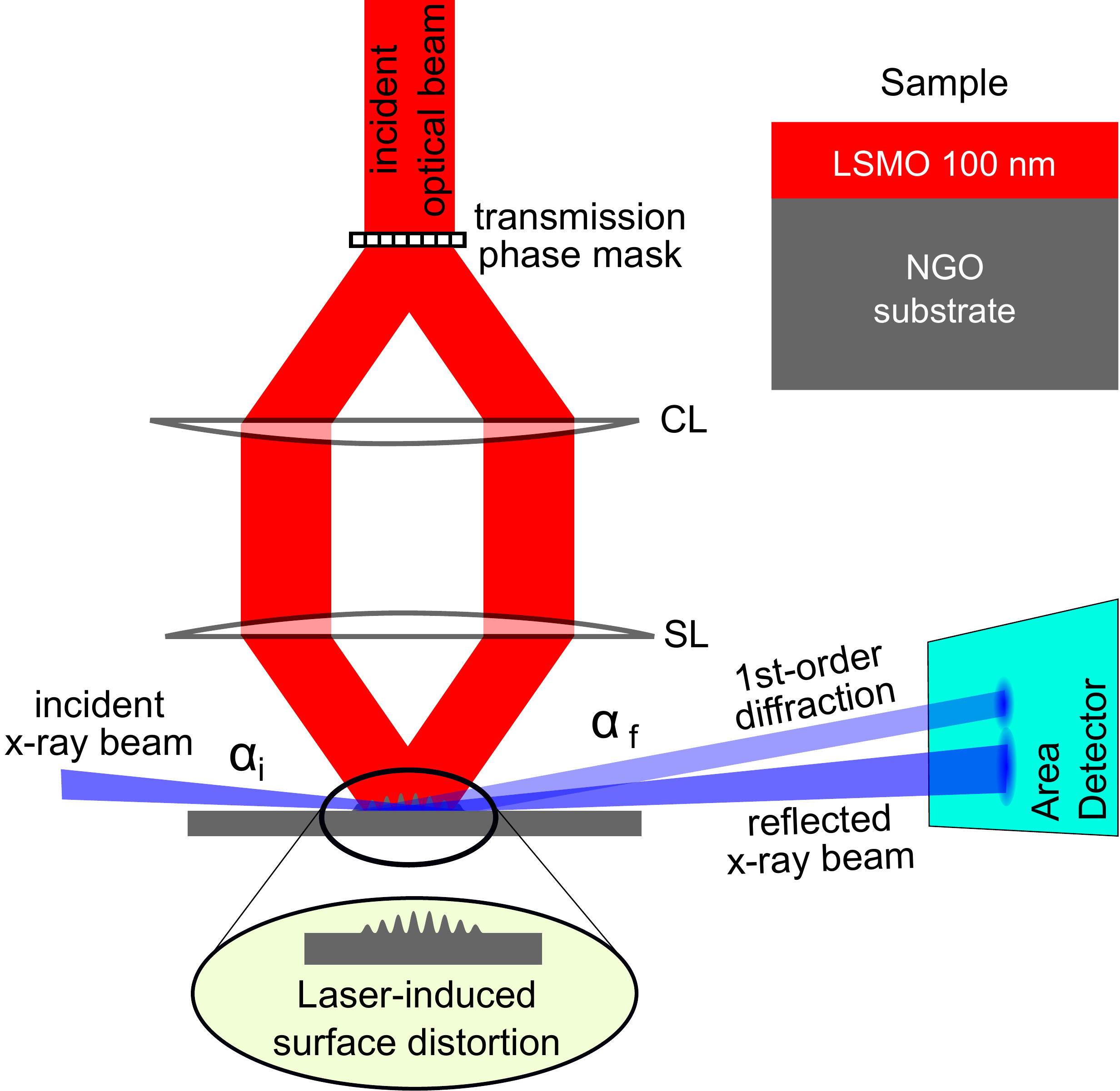}
  \caption{Experimental setup installed at the ID09 beamline at the ESRF: Transient optical gratings are generated by splitting the output of a femtosecond laser system using a transmission phase mask and combining $+1$ and $-1$ order of the optical beam on the sample surface with a cylindrical (CL) and spherical (SL) lens in 4$f$-geometry. The surface is probed with 70\,ps x-ray pulse impinging the sample at grazing incidence angle $\alpha_\text{i}$. The specular beam is reflected at the exit angle $\alpha_\text{f}$ and the first order diffraction at $\alpha_\text{f}+\delta\alpha$. We measure TR-XRR in a pump-probe scheme. A schematics of the LSMO/NGO sample structure is shown at the top.}
  \label{fig:ESRFsetup}
\end{figure}
The optical excitation pulses are coupled into the transient grating (TG) setup shown in Fig.~\ref{fig:ESRFsetup} to produce $+1$ and $-1$ diffraction order from a series of transmission phase gratings with various spatial periods $\Lambda '$. Both diffracted beams are imaged onto the sample surface using a cylindrical (CL) and spherical (SL) lens in 4$f$-geometry with focal length $f_\textrm{CL}=75$\,mm and $f_\textrm{SL}=150$\,mm, respectively. Interference of both beams at the sample surface results in a spatial light intensity distribution with spatial period $\Lambda = \Lambda' f_\textrm{SL} / f_\textrm{CL}$ which is determined by the phase grating period and the optical magnification of the setup. We define the associated wavevector $q_{\parallel}=2\pi/\Lambda$ which lies within both the sample surface and the x-ray diffraction plane. The optical setup results in a laser profile envelope at the sample surface with full-width at half maximum of $30$\,\textmu m and $4$\,mm.

The sample is probed by 70\,ps x-ray pulses with a photon energy of 15\,keV ($\lambda_{\text{x-ray}}=0.8266$\,\AA) which impinge the surface with a wavevector $\vec{k}_\text{i}$ at an incidence angle of $\alpha_\text{i}=0.15^{\circ}$, i.e., below the critical angle of total reflection. The x-ray footprint on the sample in our experiment was $\text{10\,\textmu m} \times \text{1\,mm}$, thus assuring overlap in a homogeneously excited sample area. The area detector image shows two pronounced peaks. The first peak originates from specular reflection of the incident beam at the sample surface $\vec{k}_\text{f} = \vec{k}_\text{i} + \vec{q}_{\perp}$, where $\vec{q}_{\perp} = 2\vec{k}_\text{i}\sin(\alpha_\text{i}) $ is the recoil momentum due to total reflection. The second peak is offset by an angle $\delta\alpha$ which results from the momentum transfer $\vec{q}_{\parallel}$ according to the laser-induced surface distortion. Hence, we call this peak the first-order diffraction from the laser-induced transient surface grating. In the following we investigate the temporal evolution of this first order peak.

The investigated sample consists of 100\,nm Lanthanum Strontium Manganate (LSMO) on a Neodym Gallate (NGO) substrate. It was grown by pulsed laser deposition\cite{Sell2014}. The substrate is transparent at the wavelength of the excitation laser. Hence, the optical pump pulses are absorbed exclusively in the metallic LSMO film.

\begin{figure}
  \centering
  \includegraphics[width = 0.5\textwidth]{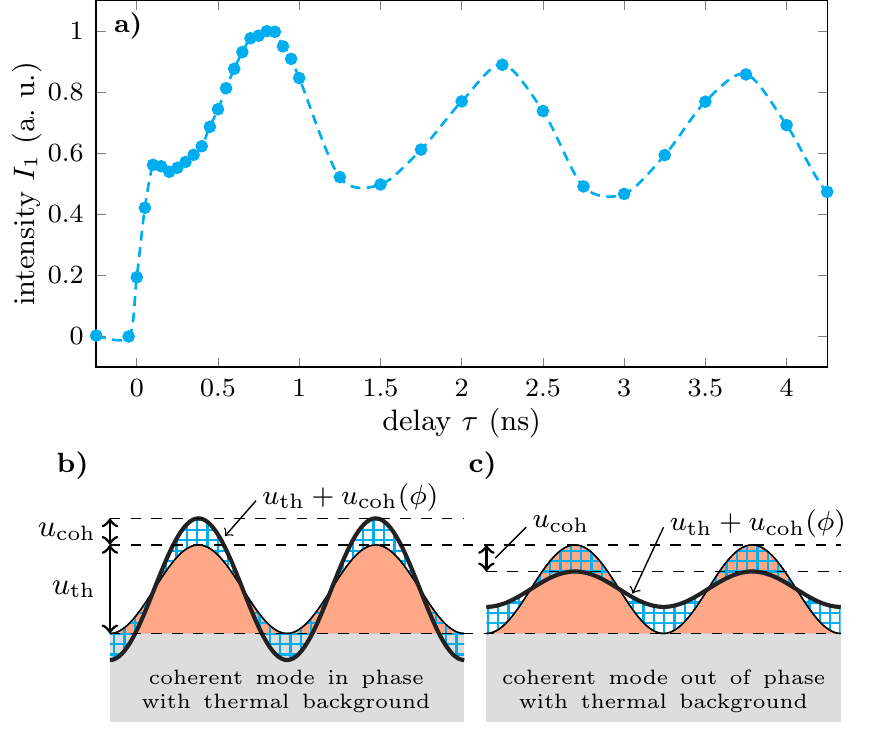}
  \caption{a) TR-XRR measurement of laser-generated transient surface deformations of a LSMO/NGO sample excited with 28\,mJ/cm$^2$. The dotted line is a guide to the eye. b) and c) Visualization of coherent and incoherent surface dynamics: the surface amplitude is modulated by constructive and destructive interference, respectively, of the periodic thermal grating and the propagating acoustic modes.}
  \label{fig:ExpData}
\end{figure}

Experimental data of the LSMO/NGO sample for an absorbed pump fluence of 28\,mJ/cm$^2$ are shown in Fig.~\ref{fig:ExpData}a). The plot depicts the intensity change of the +1st-order diffraction vs.\ pump-probe delay $(I(\tau)-I_{0})/I_{0} = \Delta I/I_{0}$. Upon optical excitation, we observe an instantaneous rise of the diffracted intensity within the temporal resolution of the experiment. The initial rise is followed by an slight intensity decay which lasts for approximately 150\,ps. The decay is followed by a signal increase which peaks at a pump-probe delay of approximately 800\,ps and subsequently oscillates around an intensity offset with constant amplitude.

The time dependence of a similar TR-XRR measurement on a different sample was recently discussed in detail\cite{Sand2017a}. Briefly, the time-resolved data can be identified to be due to a periodic surface excursion with time-dependent amplitude $u(\tau)$ described by
\begin{align}
\begin{split}
  u(\tau) &= u_{\rm th} \cdot e^{-\alpha_{\rm th}\tau} \\ 
  &\quad  + u_{\rm SAW}\cdot\cos(\omega_{\rm SAW} \tau+\varphi_{\rm SAW}) \\ 
  &\quad  + u_{\rm SSLW}\cdot\cos(\omega_{\rm SSLW} \tau + \varphi_{\rm SSLW})\cdot  e^{-\alpha_{\rm SSLW}\tau} .
\end{split}
\label{equ:SurfDistortion-1}
\end{align}
Absorption of the ultrashort light pulse in the sample results in two fundamentally different processes. First, the sample is heated locally in the excitation area, which results in a periodic thermal expansion of the surface with amplitude $u_{\rm th}$\cite{Reinhardt2016a}. In addition, the impulsive optical excitation launches coherent strain waves which propagate parallel and perpendicular to the sample surface\cite{boja2013a,herz2012c,shay2013a} and consist of two independent modes with surface displacement amplitudes $u_{\rm SAW/SSLW}$, frequencies $\omega_{\rm SAW/SSLW}$ and phases $\varphi_{\rm SAW/SSLW}$, respectively. The thermal grating decays on a timescale $1/\alpha_{\rm th} \approx 100$\,ns by in-plane thermal diffusion, a process which is much slower than the measurement range in our experiment. The SSLW mode is strongly damped with decay constant $\alpha_{\rm SSLW}$ whereas the SAW mode exhibits no decay within our measurement window.
%
A visualization of this decomposition is depicted in Fig.~\ref{fig:ExpData}b) and c). Fig.~\ref{fig:ExpData}b) depicts constructive spatial interference of the coherent modes with the thermal grating. Fig.~\ref{fig:ExpData}c) shows a situation where the thermal grating and the coherent modes are spatially in opposite phase. Hence, both excitations interfere destructively. The interplay of coherent and incoherent excitations can be exploited for spatiotemporal control of the surface excursion\cite{Sand2017a}.

\begin{figure*}
  \centering
  \includegraphics[width = \textwidth]{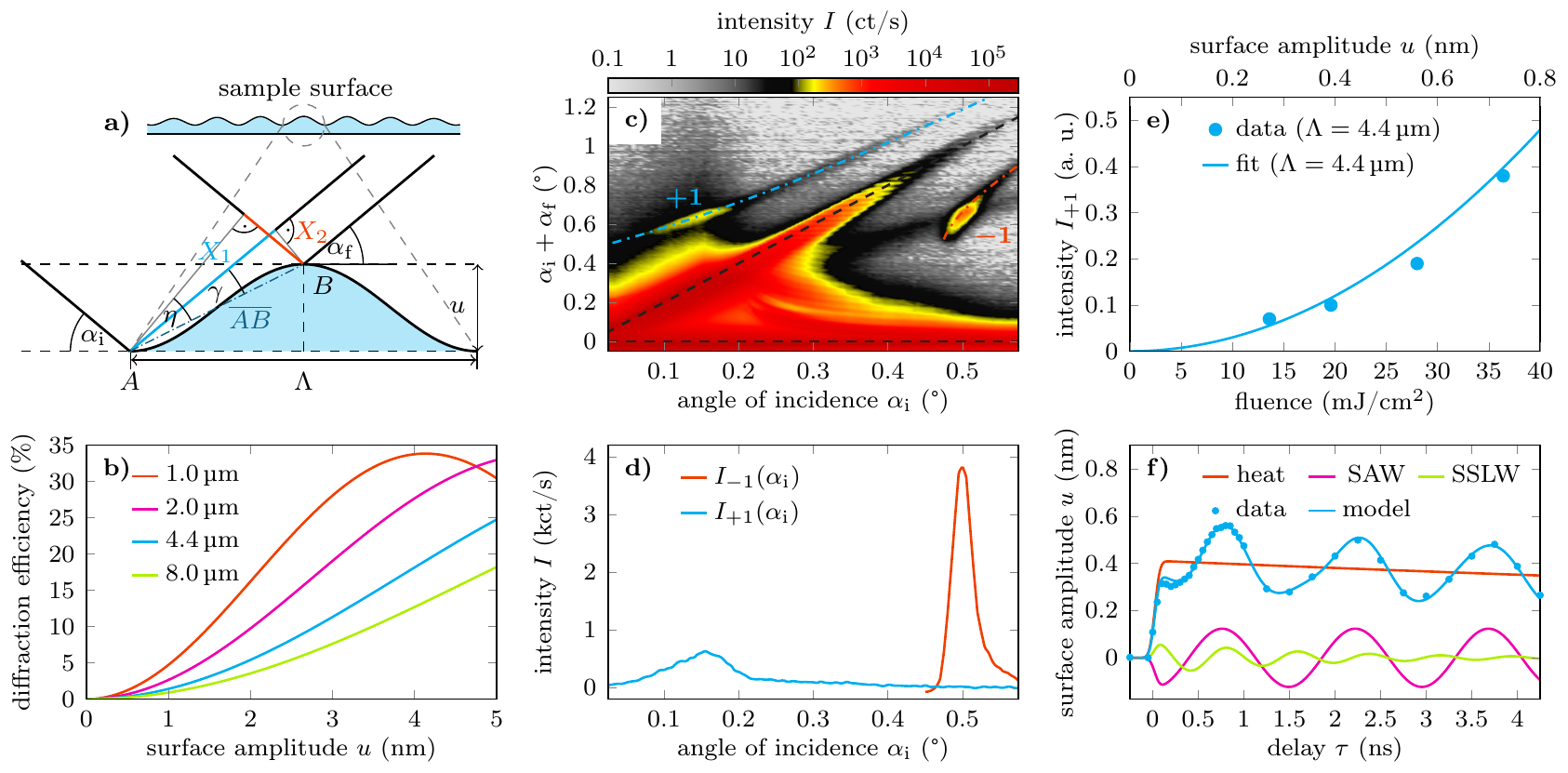}
  \caption{a) Schematic for the diffraction model given by eqs.~(\ref{equ:SurfModel-b} - \ref{equ:SurfModel-e}). b) Relative diffraction efficiency vs.\ surface excursion for grating periods of $\Lambda=8.0\,\text{\textmu m}$ (green), 4.4\,\textmu m (blue), 2.0\,\textmu m (magenta) and 1.0\,\textmu m (orange). c) Angle-resolved diffracted intensity vs.\ incidence angle $\alpha_\text{i}$. The specular reflection (black dashed line) and $+1$st and $-1$st order diffraction from the surface grating (blue and red dashed lines) are indicated. d) Intensity along the $+1$st and $-1$st diffraction order, i.e., along the colored dashed lines in c). e) Diffracted intensity (symbols) vs.\ absorbed pump fluence (bottom) and surface amplitude (top). The solid line shows a quadratic dependence of the diffracted intensity as expected from eq.~\eqref{equ:DiffractionModel}. f) Decomposition of the diffraction data for an absorbed pump fluence of 28\,mJ/cm$^2$. The amplitude of the individual components are given in absolute scale.}
  \label{fig:DiffModel}
\end{figure*}
Here we explicitly analyze the TR-XRR probing mechanism to derive a diffraction model which relates the diffracted intensity $I(\tau)$ with the amplitude $u$ of the surface excursion. A spatial period of the distorted sample is shown in Fig.~\ref{fig:DiffModel}a). An x-ray beam impinges the sample at an incidence angle $\alpha_\text{i}$ in the bottom (point $A$) and on the top (point $B$) of the distortion and is reflected with an exit angle $\alpha_\text{f}$. After reflection from $A$, the beam travels an additional path length $X_{1}$, while the other beam travels an additional path $X_{2}$ before reflection from point $B$. The total path difference results in a relative phase of both beams of $\Delta\phi=\frac{2\pi}{\lambda_\text{x-ray}}\cdot(X_{2}-X_{1})$. $\Delta\phi$ can be calculated using the following set of equations:
\begin{eqnarray}
  X_{1}&=& \sqrt{u^{2}+(\Lambda/2)^{2}}\cdot\cos(\gamma)  \label{equ:SurfModel-b} \\
  X_{2} &=& \sqrt{u^{2}+(\Lambda/2)^{2}}\cdot\sin(\eta) \label{equ:SurfModel-c} \\
  \gamma &=& \alpha_\text{f}-\tan^{-1}(2u/\Lambda)  \label{equ:SurfModel-d} \\
  \eta &=& \pi/2-\alpha_\text{i}-\tan^{-1}(2u/\Lambda) \label{equ:SurfModel-e}
\end{eqnarray}
and the grating equation for constructive interference:
\begin{eqnarray}
  n \lambda_\text{x-ray} &=& \Lambda \left( \cos(\alpha_\text{f}) - \cos(\alpha_\text{i}) \right) \label{equ:SurfModel-f}
\end{eqnarray}
From kinematical theory of surface diffraction \cite{Madsen2005a, Nicolas2014a} we find the following expression for the diffraction intensity of $n$-th order from a periodically distorted surface for incidence angles below the critical angle $\alpha_\text{i}<\alpha_\text{c}$, i.e.\ from a pure phase grating,
\begin{eqnarray}
 I_n & = &\left|\frac{1}{r_{0}}\int_{\parallel}e^{-i\left(n q_{\parallel} r_{\parallel}+\frac{\Delta\varphi}{2}\sin(\frac{2\pi}{\Lambda}r_{\parallel})\right)}\text{d}r_{\parallel}\right|^{2} \label{equ:DiffInt1}\\
 & = & \left|J_n\left(\frac{\Delta\varphi}{2}\right)\right|^{2} \label{equ:DiffInt2}
\end{eqnarray}
where $r_\parallel$ is the spatial coordinate along the surface grating, $r_{0}$ is a normalization constant and $J_n$ is the $n$-th Bessel function. The argument of the Bessel function is the modulation of the phase difference due to variation of the grating surface amplitude $\Delta\varphi=\Delta\phi-n\pi$, where $n\pi$ is the phase shift due to $n$-th order diffraction. For all practical purposes we can assume that the surface amplitude is much smaller than the period of the surface grating, i.e.\ $u\ll\Lambda/2$ and therefore $\sqrt{u^{2}+(\Lambda/2)^{2}}\approx\frac{\Lambda}{2}(1+\frac{2u^2}{\Lambda^{2}})$ and $\tan^{-1}(\frac{2u}{\Lambda}) \simeq\tan(\frac{2u}{\Lambda})\simeq\frac{2u}{\Lambda}$. For grazing incidence $\alpha_\text{i}\leq\alpha_\text{c}$ the phase difference $\Delta\varphi$ is approximately given by
\begin{align}
\Delta\varphi= -2\pi\frac{u}{\lambda_{\text{x-ray}}}\alpha_{\text{i}}\Biggl[1+\sqrt{1+\frac{2n\lambda_{\text{x-ray}}}{\Lambda\alpha_{\text{i}}^{2}}}\,\Biggr]
\label{equ:DiffractionModel}
\end{align}
Results of the diffraction model laid out by eqs.~(\ref{equ:SurfModel-b}-\ref{equ:DiffInt2}) are presented in Fig.~\ref{fig:DiffModel}b). We plot the normalized diffracted first-order intensity $I_1$ vs.\ the surface excursion $u$ for spatial grating periods $\Lambda=8.0\,\text{\textmu m}$ (pink), 4.4\,\textmu m (blue), 2.0\,\textmu m (green) and 1.0\,\textmu m (red). The diffraction efficiency increases with increasing surface excursion and with decreasing spatial period $\Lambda$. The maximum diffraction efficiency is 33\%, i.e.\ the maximum of the Bessel function shown in Fig.~\ref{fig:DiffModel}b). Experimental data is shown in Fig.~\ref{fig:DiffModel}c) which depicts the diffracted intensity from the sample vs.\ incidence angle $\alpha_\text{i}$. The specular reflection and the $\pm1$st-order diffraction are marked by dashed lines. Integrated intensity of the $+1$st (blue) and $-1$st diffraction order are shown in Fig.~\ref{fig:DiffModel}d). The integration was performed along the colored dashed lines in panel~\ref{fig:DiffModel}c).

The fluence dependence of the $+1$st diffraction order intensity from a laser-generated surface grating with spatial period \mbox{$\Lambda$ = 4.4\,\textmu m} is depicted in Fig.~\ref{fig:DiffModel}e). The symbols indicate the measured maximum diffracted intensity at 800\,ps time delay vs.\ absorbed pump fluence. The total surface excursion at this time delay is the sum of a thermal grating and of coherent sound waves with an out-of-plane polarization component (see Fig.~\ref{fig:ExpData}b)). Using recent time-resolved x-ray diffraction data from a similar LSMO sample\cite{boja2012b}, a calibration factor for the laser-generated layer strain under the given circumstances can be estimated to be approx. 0.02\% per mJ/cm$^{2}$. \footnote{Note that the calibration factor given in Ref.~\onlinecite{boja2012b} refers to the incident fluence instead of the absorbed fluence and accounts only for the thermal component. At 800\,ps the in-phase coherent modes require an enhanced calibration factor of 0.02\% per mJ/cm$^{2}$ which is applied in Fig.~\ref{fig:DiffModel}e)} Taking into account the LSMO layer thickness of 100\,nm, the experimental fluence can thus be converted to a total surface excursion which is given at the top abscissa of Fig.~\ref{fig:DiffModel}e). The solid line shows results from our diffraction model presented in eqs.~(\ref{equ:SurfModel-b} - \ref{equ:SurfModel-e}). The experimental data shows the expected quadratic dependence as derived from eq.~\eqref{equ:DiffractionModel}.

The time-resolved surface dynamics upon transient grating excitation is shown in Fig.~\ref{fig:DiffModel}f). By taking the square root of the diffracted intensity, i.e.\ data shown in Fig.~\ref{fig:ExpData}a), we depict the surface excursion on an absolute length scale. Experimental data (symbols) are decomposed into a slowly decaying thermal grating (red), a Rayleigh-like SAW mode (magenta) and a SSLW-mode (green), respectively. The solid blue line shows the time-dependent surface dynamics given as described by eq.~\eqref{equ:SurfDistortion-1}, showing excellent agreement with the experimental curve.

In conclusion we measure the absolute amplitude of the surface excursion of a laser-induced transient grating on a solid surface by time-resolved x-ray reflectivity. Ultrafast optical excitation generates incoherent thermal surface distortions and coherent acoustic surface waves. The measured dynamics at the surface allow for a decomposition of the surface amplitude in a thermal background and two coherent acoustic modes: a Rayleigh-like surface acoustic wave and a surface skimming longitudinal wave. Our new method can be applied to decompose coherent and incoherent surface dynamics with sub-\AA{} precision and with a temporal resolution better than 100\,ps.

The experiments were performed at the beamline ID09 of the European Synchrotron Radiation Facility (ESRF), Grenoble, France. We are grateful to Gemma Newby, Martin Pedersen and Micheal Wulff for providing assistance in using beamline ID09. We also thank Jutta Schwarzkopf for preparation of the sample. We acknowledge financial support from BMBF via 05K16GU3, from {\it Strat\'{e}gie Internationale} "NNN-Telecom" de la R\'{e}gion Pays de La Loire, ANR-DFG "PPMI-NANO" (ANR-15-CE24-0032 \& DFG SE2443/2) and from DFG via BA2281/8-1.

%

\end{document}